\documentclass{article}
\usepackage{frascatiphys}
\begin{document}        
\title{IDENTIFICATION OF SHOWERS WITH THE CORE OUTSIDE THE ARGO-YBJ DETECTOR}
\author{
Giuseppe Di Sciascio, Elvira Rossi on behalf of the ARGO-YBJ Collaboration  \\
{\em Dip. di Fisica Universit\`a di Napoli and INFN, sez. di Napoli, Italy} \\
}
\maketitle
\baselineskip=11.6pt
\begin{abstract}
In this paper we present a procedure able to identify and reject showers with the core outside 
the ARGO-YBJ carpet boundaries. The efficiency of this method is investigated for different 
primary energies and fiducial areas. A comparison of the results for gamma and proton induced 
showers is also presented.
\end{abstract}
\baselineskip=14pt

\section{Introduction}
The ARGO-YBJ detector, currently under construction at the Yangbajing 
Laboratory (P.R. China, 4300 m a.s.l.), is a full coverage array of dimensions 
$\sim 74\times 78~m^2$ realized with a single layer of RPCs. The area 
surrounding this central detector ({\it carpet}), up to $\sim 100 \times 110~m^2$, 
is partially ($\sim 50 \%$) instrumented with other RPCs.
The basic element is the logical {\it pad} ($56\times 62~cm^2$) which defines 
the time and space granularity of the detector. The layout of the detector is shown in 
Fig. \ref{argo}.
The detector is subdivided in 6 $\times$ 2 RPC units (Clusters, the rectangles of 
Fig. \ref{argo}).

Showers of sufficiently large size will trigger the detector even if their core is located 
outside its boundaries. 
The corresponding core positions are generally reconstructed not only
near the carpet edges but also well inside them.
As a consequence, sofisticated algorithms to reduce the contamination of external events 
are needed.
The goal is to identify and reject a large fraction of external events before 
exploiting any reconstruction algorithm only by using some suitable parameters.

The rejection of external events is important because a large difference between the true and 
the reconstructed shower core position may lead to a systematic miscalculation of some shower 
characteristics, such as the shower size. 
Moreover, an accurate determination of the shower core position for selected 
internal events is important to reconstruct the primary direction using conical fits to the 
shower front, improving the detector angular resolution or to performe an efficient gamma/hadron 
discrimination.

In this paper we present a reconstruction procedure able to identify and reject a large fraction
of showers with core outside the ARGO-YBJ detector. 
The efficiency of this procedure is investigated both for gamma and proton induced showers.

\section{Identification of external events}
\label{cut}

To perform these calculations we have simulated, via the Corsika code\cite{corsika}, $\gamma$-induced showers with a Crab-like spectrum ($\sim~E^{-2.5}$) and 
proton events with $\sim~E^{-2.75}$, both ranging from 100 GeV to 50 TeV. The $\gamma$-rays have been simulated for different zenith angles ($<40^{\circ}$), following the daily path of the source in the sky. 
The detector response has been simulated via a GEANT3-based code.
The core positions have been randomly sampled in an area, 
energy-dependent, large up to $800 \times 800~m^2$ centered on the detector. 

\begin{figure}[t]
\begin{minipage}[t]{.48\linewidth}
   \vspace{5.8cm}
\includegraphics{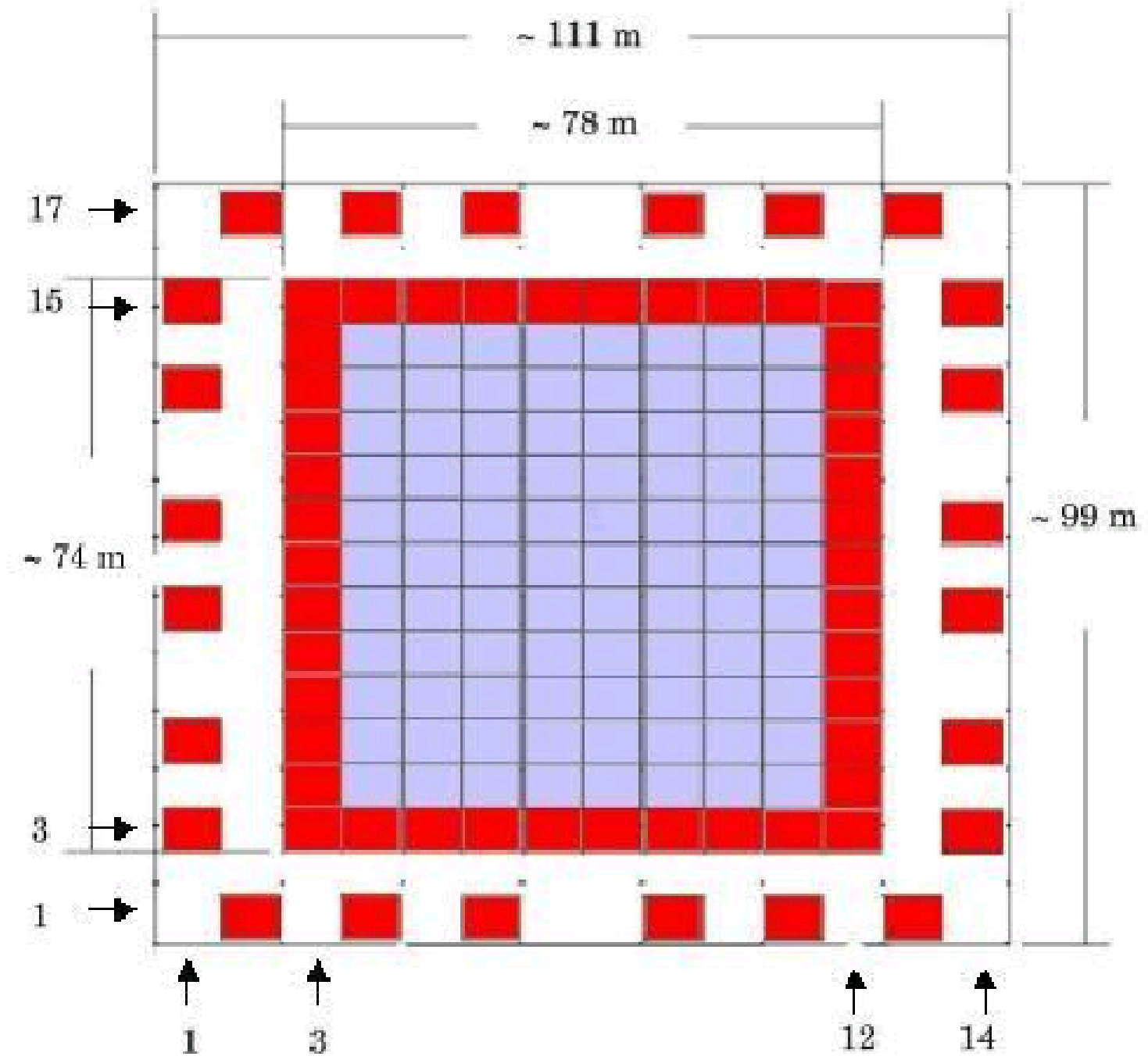}
 \caption{\it The ARGO-YBJ detector.
    \label{argo} }
 \end{minipage}\hfill     
\begin{minipage}[t]{.48\linewidth}
   \vspace{5.8cm}
    \includegraphics{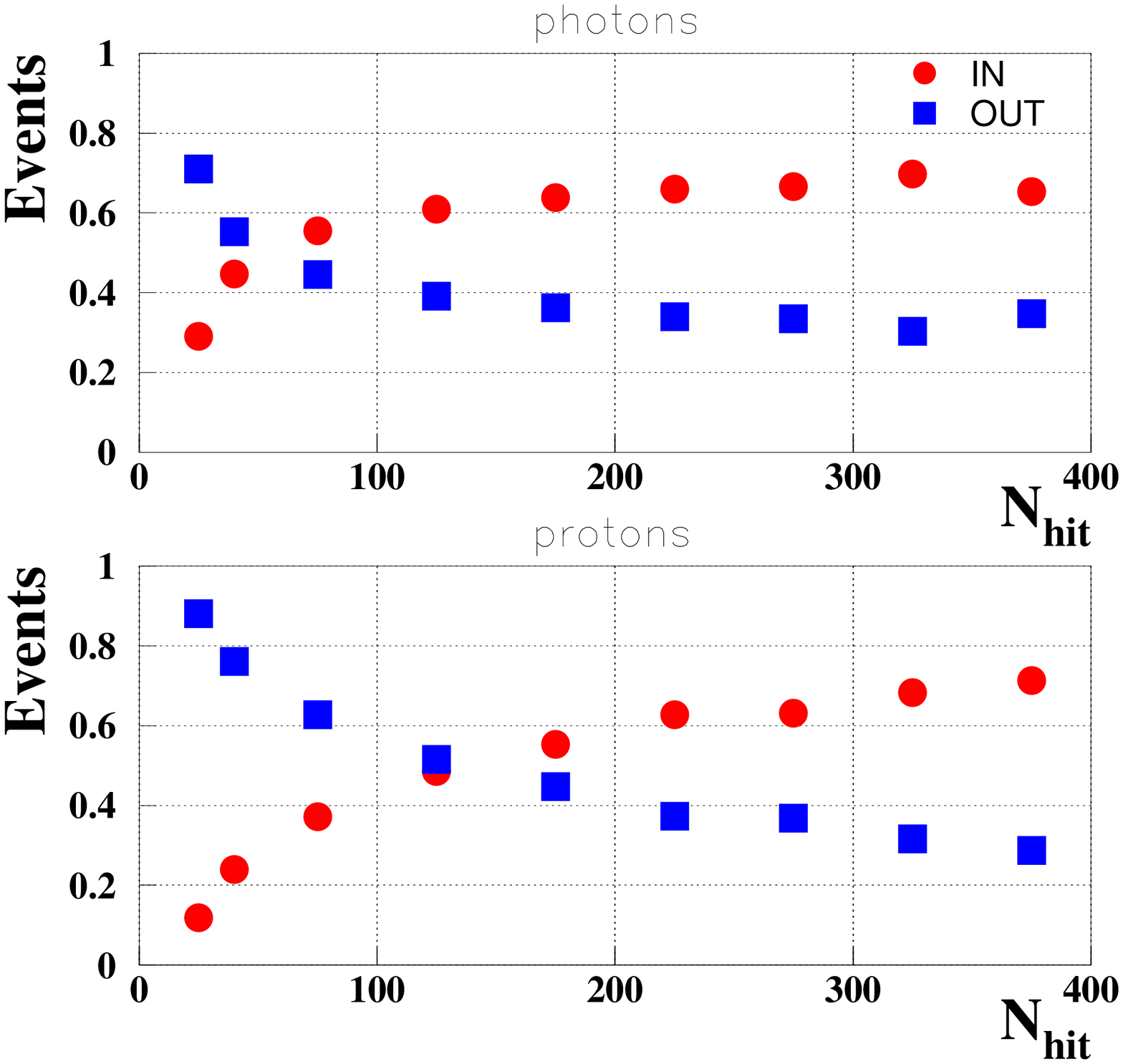}
    \caption{\it Fraction of events with the core IN (OUT) 
$A_{fid}=80 \times 80~m^2$.
 \label{nevents} }
 \end{minipage}\hfill        
\end{figure}

In Fig. \ref{nevents} we show the fraction of showers with the core truly internal (external)
to a fiducial area approximately equal to the carpet dimensions ($A_{fid}=80 \times 80~m^2$) as a 
function of the number of pads fired on the ARGO-YBJ carpet. 
The upper plot refers to $\gamma$-induced showers, the lower to proton-induced events.
As expected, for low pad multiplicities the events are mainly external:
about $40 \%$ of the $\gamma$-induced showers with a multiplicity of 100 - 150 fired pads 
are external to $A_{fid}$.
But also for higher multiplicities the percentage of external events is consistent (about 
$30 \%$).

Various parameters based on particle density or time information are under investigation 
to identify showers with core position outside a given fiducial area\cite{elly}. 
In this paper we discuss the performance of the following ones:
\begin{itemize}
\item Position of the cluster with the highest particle density.
\item Position of the cluster row/column with the highest particle density.
\item Mean distance $R_p$ of all fired pads to the reconstructed shower core position.
\end{itemize}

\begin{figure}[t]
\begin{minipage}[t]{.48\linewidth}
   \vspace{5.8cm}
    \includegraphics{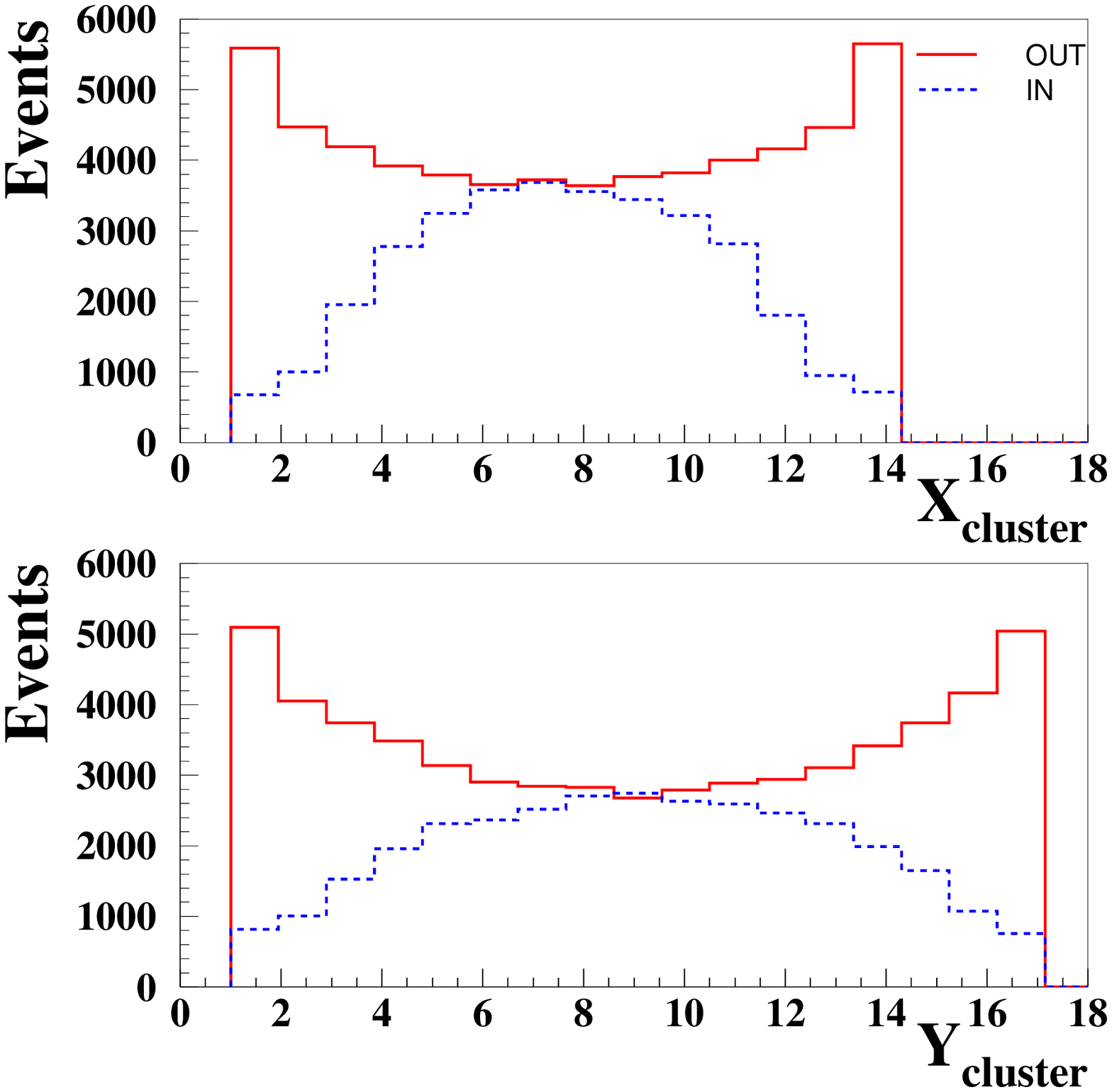}
    \caption{\it Coordinate distributions of the cluster with the highest particle density 
for $\gamma$-induced events with pad multiplicity $N_{hit} > 100$. 
 \label{clusmax} }
 \end{minipage}\hfill
\begin{minipage}[t]{.48\linewidth}
   \vspace{5.8cm}
    \includegraphics{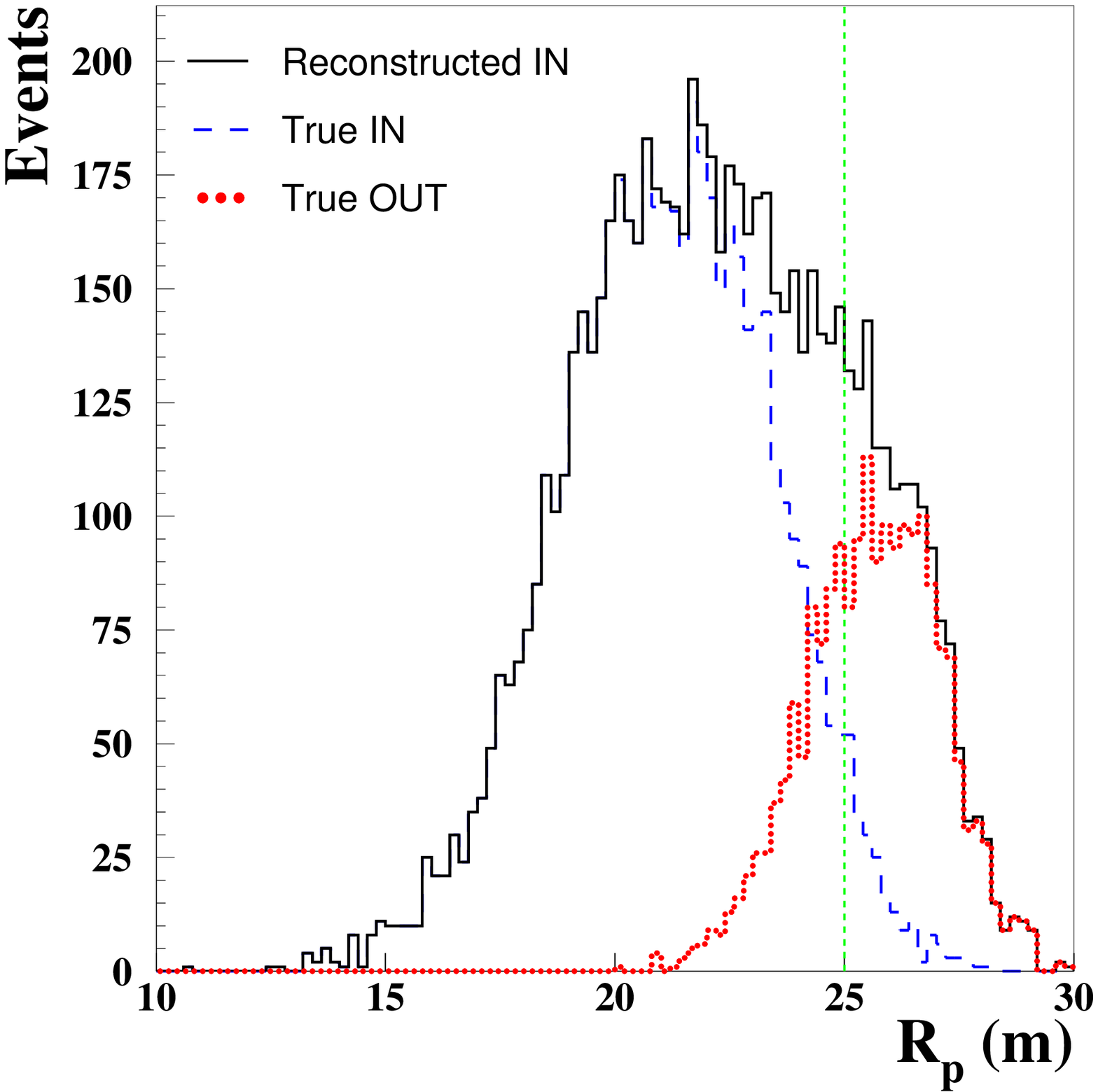}
    \caption{\it Distributions of the parameter $R_p$ for $\gamma$-induced events 
with $N_{hit} > 100$. 
 \label{rppad} }
\end{minipage}\hfill
\end{figure}

As an example, in Fig. \ref{clusmax} we show the distributions of the positions of the cluster 
with the highest particle density for $\gamma$-induced showers which fire more than 100 pads
on the central carpet.
In the plots we compare the events with the core truly external to a 80 $\times$ 80 
m$^2$ fiducial area (solid histograms) and the truly internal ones (dashed histograms).
To investigate the discrimination power of this particular parameter we have simulated
a detector completely instrumented up to $\sim$ 100 $\times$ 110 $m^2$, i.e., 
containing 14 $\times$ 17 clusters.
Therefore, the cluster coordinates run from 1 to 14 (X view) and from 1 to 17 (Y view) 
starting from the lower left corner of the carpet (see Fig. \ref{argo}). 
The clear difference between the IN and OUT showers suggests to tag as 
external the events with the highest particle density in the outer clusters and to reject them 
before exploiting any reconstruction algorithm.

The mean lateral spread of the shower can be expressed as:
\begin{equation}
R_p = \displaystyle\frac{\sum_{i=1}^{N} \,{|r_C - r_i|} \, n_i}
                                   {\sum_{i=1}^{N} \, n_i}
\end{equation}
where $N$ is the number of fired pads, $r_C$ is the reconstructed core position, $r_i$ is the
position of the $i$-th fired pad, $n_i$ is the number of detected electrons in the $i$-th pad.
The $R_p$ distribution for showers reconstructed inside a 80 $\times$ 80 m$^2$
fiducial area is shown in Fig. \ref{rppad} (solid histogram). 
The dashed line refers to truly IN events while the dotted histogram refers to OUT
showers erroneously reconstructed as internal. 
The shower cores have been calculated by means of the simple center of gravity method.
As can be seen, the parameter $R_p$ identifies quite well the events with core outside 
the carpet. Large distances between the true and the reconstructed shower axis lead to larger 
$R_p$ values. This fact offers the possibility to define a cut in $R_p$ to 
identify these events. A conservative choice is to reject showers with 
$R_p > 25~m$. 

From these studies it follows that the identification of a large fraction of external events 
can be achieved by defining a suitable fiducial area and a combination of 
cuts in the parameters discussed above.

\section{Maximum Likelihood Method ({\tt LLF}) }

Different algorithms have been investigated to reconstruct the shower core position in the
ARGO-YBJ experiment\cite{icrc01}. The most performant one is the Maximum Likelihood Method.

If $<m_i>$ is the average particle number expected on the $i$-th pad, then the probability 
of finding $N_i$ particles is
\begin{equation}
 P_i={ {<m_i>^{N_i}} \over {N_i!} }\cdot e^{-<m_i>}
\end{equation}
Therefore $P_i(0) = e^{-<m_i>}$ is the probability of finding 0 particles, and
$P_i(>0) = 1-e^{-<m_i>}$ is the probability of finding 1 or more particles.
The Likelihood Function (LF) is given by: $LF = \Pi_i P_i$.
The natural Logarithm of LF ({\tt LLF}) becomes a sum:
\begin{equation}
 LLF = ln(\Pi_i P_i)=\Sigma_i ln P_i(0) + \Sigma_j ln P_j(>0)
\end{equation}
where the index $i$ runs on the not fired pads, while the index $j$ refers to the
fired pads. Exploiting the relation $<m_i>=S_{pad}\cdot \rho_i$, we obtain
\begin{equation}
 -LLF \equiv S_{pad}\cdot \Sigma_{i} \rho_i - \Sigma_{j} ln(\rho_j) - 
N_{pad}(>0)\cdot ln(S_{pad}) 
\label{eq-llf1}
\end{equation}
where $\rho_i = f(R_i/R'_M) \cdot {{N_e} \over {R_M^{'2}}}$ is the particle density expected on
the $i$-th pad at a distance $R_i$ from the core. The lateral structure function $f(R_i/R'_M)$ 
has been calculated for the ARGO-YBJ detector:
\begin{equation}
 f \left ( {\frac {R_i}{R'_M} } \right )  = C \cdot R_i^{A-2} \cdot \left ( 1+ \frac {R_i}{R'_M}
\right ) ^{-B} 
\end{equation}
with the normalization factor defined by
\begin{equation}
 C = \frac {\Gamma(B)}{2\pi (R'_M)^{A-2} \Gamma(A)\Gamma(B-A)} 
\end{equation}
where $R'_M=R_M/3.944$, $R_M$ being the Moliere radius (133 m at YBJ altitude), 
and A = 1.826, B = 2.924,  C = 0.613 .
$S_{pad}$ is the pad area and $N_{pad}(>0)$ is the total number of pads fired by the 
shower. The minimum value of -LLF is then chosen as the best fit for the freely 
varying parameters $\{x_c,y_c,N_e\}$, being $\{x_c,y_c\}$ the core coordinates and 
$N_e$ the shower size.
In the following we will refer to this approach as to the ''{\tt LLF1} method''.

\begin{figure}[t]
\begin{minipage}[t]{.48\linewidth}
 \vspace{5.8cm}
    \includegraphics{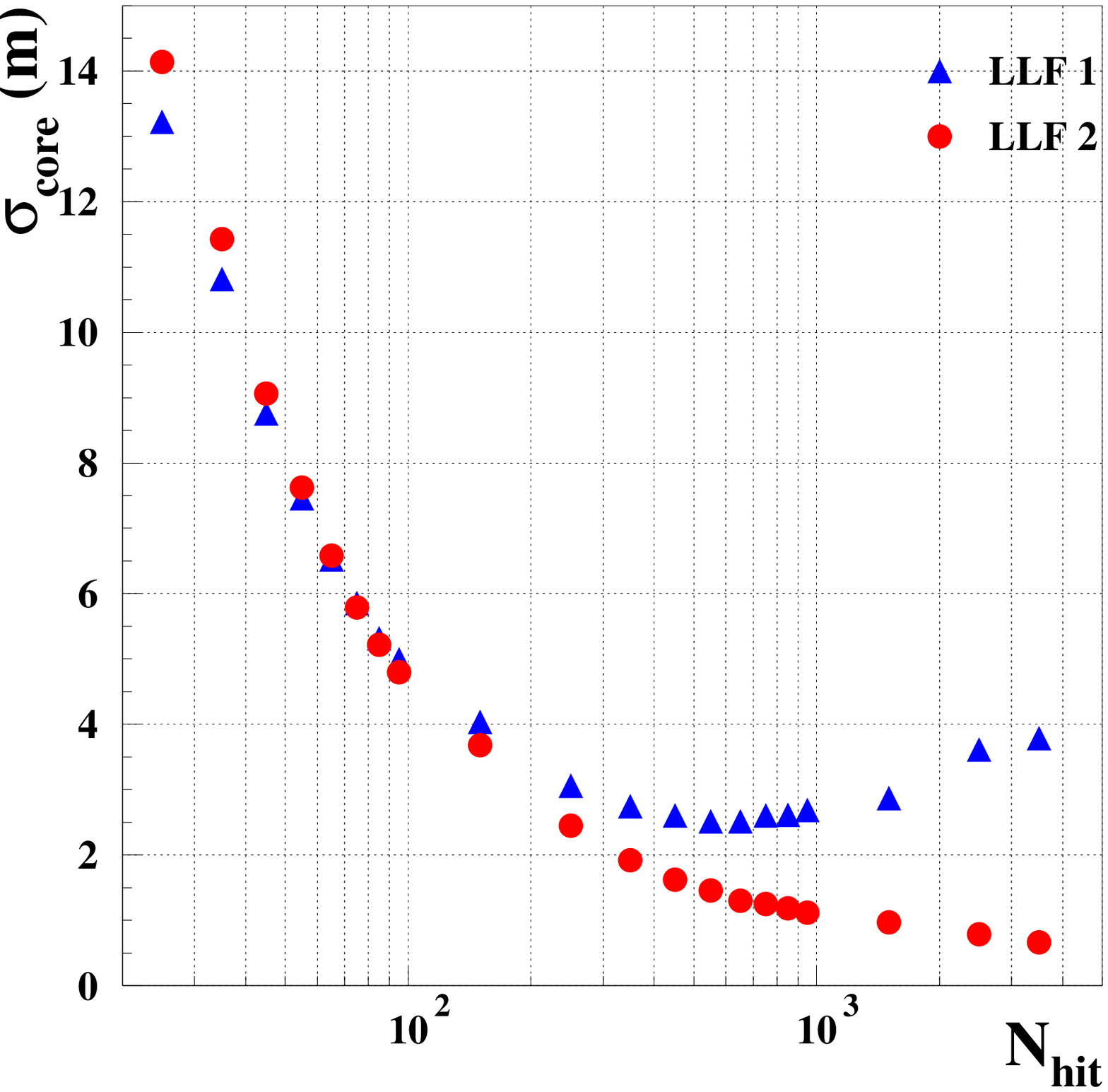}
    \caption{\it Comparison between the shower core position resolutions 
obtained with {\tt LLF1} and {\tt LLF2} methods. 
 \label{llf1_llf2} }
\end{minipage}\hfill
\begin{minipage}[t]{.48\linewidth}
\vspace{5.8cm}
    \includegraphics{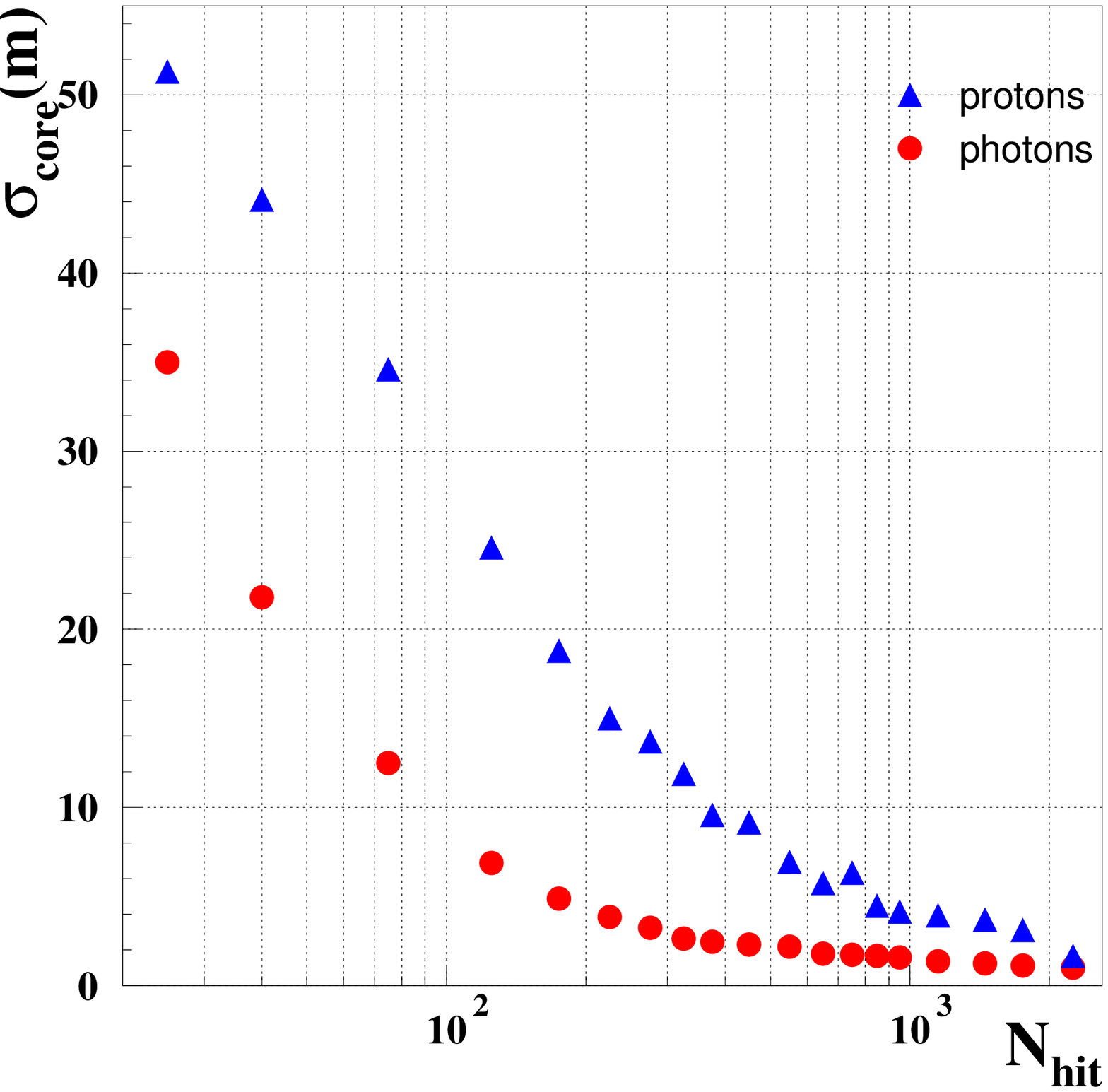}
    \caption{\it Shower core position resolutions obtained  with {\tt LLF2} 
for $\gamma$ (circles) and proton (triangles) showers.
 \label{llf2} }
\end{minipage}\hfill
\end{figure}

We point out that expression (\ref{eq-llf1}) for -LLF refers to the case of a Poisson 
distribution in which the pads are not fired with 
probability $P_i (0)$ or fired with probability $P_i (>0) = 1 - P_i (0)$. 
In our study almost always the fired pads have particle multiplicity 1, and therefore such a 
simple discrimination can be made. However, if we consider a larger area as the
whole RPC, the multiplicity can also be $>$ 1, and the proper Poisson
distribution on the fired RPCs appears more adequate. In this case the sum on fired elements is:
\begin{equation}
-\Sigma_j ln P_j (>0) = -\Sigma_j N_j ln(\rho_j) - ln(S_{RPC}) \Sigma_j N_j
 +\Sigma_j ln(N_j !) + S_{RPC} \Sigma_j \rho_j
\end{equation}
where $\rho_j$ is the particle density expected on the j-th RPC at a distance $R_j$ 
from the core, $N_j$ is the recorded particle number and $S_{RPC}$ is the RPC area. 
The shower size can be calculated via the equation
\begin{equation}
 N_e =  \frac{\Sigma_j N_j } {S_{RPC}\Sigma_j \rho_j}
\end{equation}
We call this calculation the ''{\tt LLF2} method''.
As a consequence, we expect that the differences between {\tt LLF1} and {\tt LLF2}
increase, for a fixed area,with the particle density.
In Fig. \ref{llf1_llf2} we compare the shower core position resolution 
$\sigma_{core}=\sqrt{\sigma_x^2 + \sigma_y^2}$ calculated by applying 
the {\tt LLF1} and {\tt LLF2} methods on the RPCs for $\gamma$-induced showers with the core
randomly sampled inside a $80 \times 80~m^2$ area. 
As expected, the resolution worsens with the multiplicity if the {\tt LLF1} approach is applied 
when the number of particles hitting the RPC is $>~1$.
We note that for very low multiplicities ($N_{hit} < 80$) method {\tt LLF1} is more 
performant than {\tt LLF2}. Indeed, the algorithm based on the RPC occupancy ({\tt LLF1})
provides a better representation of the hit distribution in very poor showers.
For very high multiplicities ($N_{hit} > 10^3$) the shower core position 
is determined by {\tt LLF2} with an uncertainty $<1~m$.

In Fig. \ref{llf2} $\sigma_{core}$ obtained with the {\tt LLF2} method is shown as a function of 
multiplicity for $\gamma$ and proton induced showers with core randomly sampled in an area,
energy-dependent, large up to $800 \times 800~m^2$. 
The fiducial area is $80 \times 80~m^2$. The increase of the sampling area worsens the 
performance of the method because of the contamination by external events 
(see Fig. \ref{nevents}). 
This effect can be limited exploiting the cuts described in section \ref{cut}.
As an example, in Fig. \ref{llf2_rp} we compare the shower core position resolutions 
for $\gamma$ and proton showers after the following selection procedure:
(1) Rejection of the events whose highest density clusters are on the guard ring 
($X =\{ 1, 14\}$; $Y =\{ 1, 17\}$) or on the boundaries of the central carpet ($X = \{3, 12\}$; $Y = \{3, 15\}$);
(2) Rejection of the events whose highest total density rows or columns are respectively 
in positions $\{1,3,15,17\}$ or $\{1,3,12,14\}$;
(3) Reconstruction of core coordinates $\{ X_c,Y_c \}$ using {\tt LLF2};
(4) Further rejection of events with $R_p > 25$ m.

For events with a multiplicity of $100-150$ hits $\sigma_{core}$ improves 
from $6.9$ to $4.4~m$ for $\gamma$-induced showers and from $24.6$ to $21.3~m$ for 
proton events.   
For any given multiplicity the shower core position resolution is better for $\gamma$ than 
for proton induced showers, due to the larger lateral particles spread in the latter.

\section{Results }

\begin{figure}[t]
\begin{minipage}[t]{.48\linewidth}
 \vspace{5.8cm}
    \includegraphics{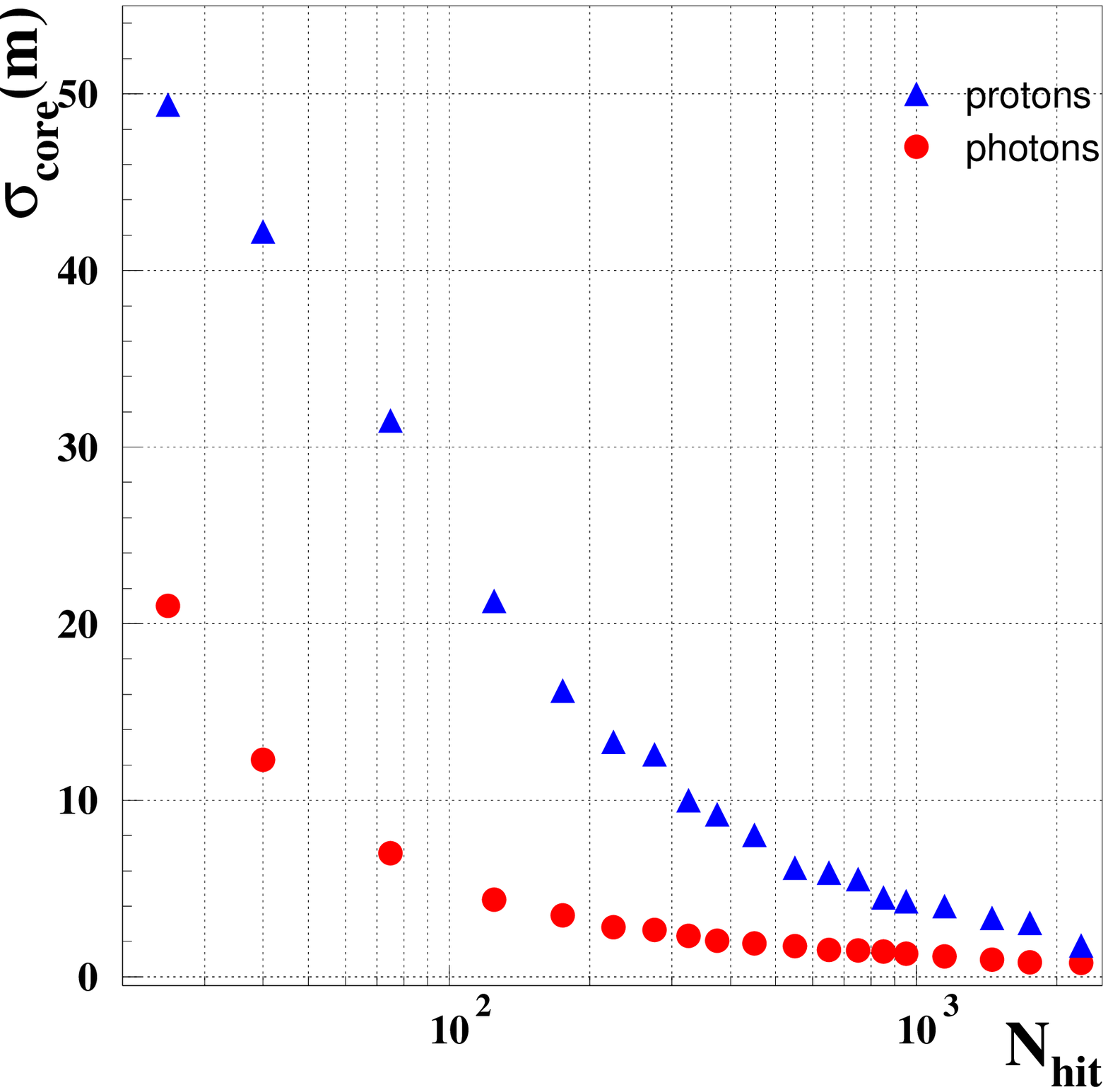}
    \caption{\it Comparison between the shower core position resolutions for 
$\gamma$ (circles) and proton (triangles) events after the selection 
procedure (1) - (4). 
 \label{llf2_rp} }
\end{minipage}\hfill
\begin{minipage}[t]{.48\linewidth}
 \vspace{5.8cm}
\includegraphics{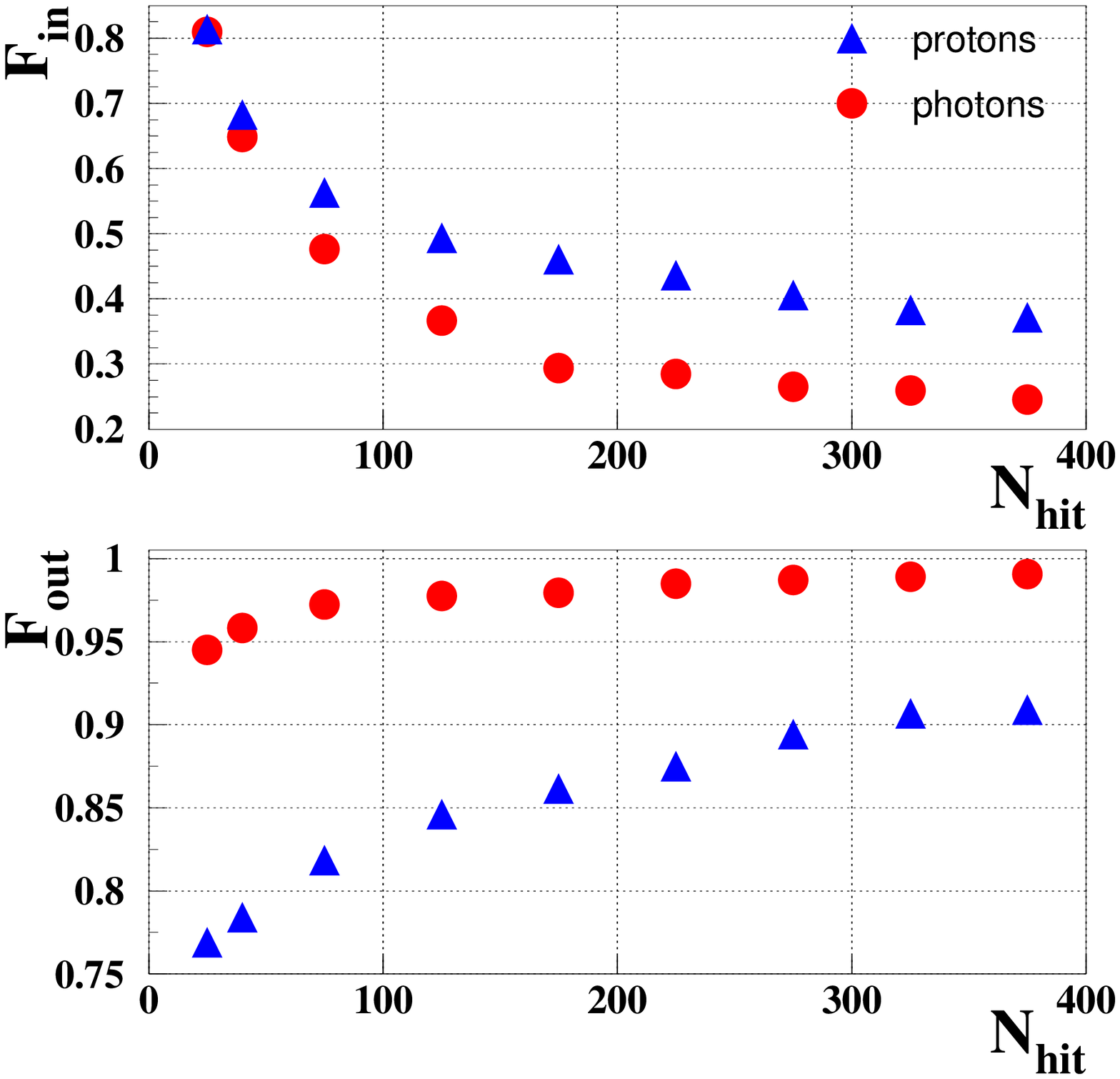}
 \caption{\it Fraction of truly internal ($F_{in}$) and external ($F_{out}$) events rejected by 
the selection procedure (1) - (4).
 \label{finfout_4} }
\end{minipage}\hfill
\end{figure}

The procedure (1) - (4) is one of the possible procedures to reject external events in the
ARGO-YBJ detector.
In Fig. \ref{finfout_4} the fraction of rejected events (internal and external to 
$A_{fid}=80 \times 80~m^2$, respectively) is reported for $\gamma$ and proton-induced showers.
As can be seen, this algorithm is able to identify and reject a large fraction of external events:
for photons this percentage is always larger than $95 \%$.
For low multiplicities ($N_{hit} < 150$) a significative fraction of internal events is 
erroneously rejected, especially in proton-induced showers.

\begin{figure}[t]
\begin{minipage}[t]{.48\linewidth}
\vspace{5.8cm}
    \includegraphics{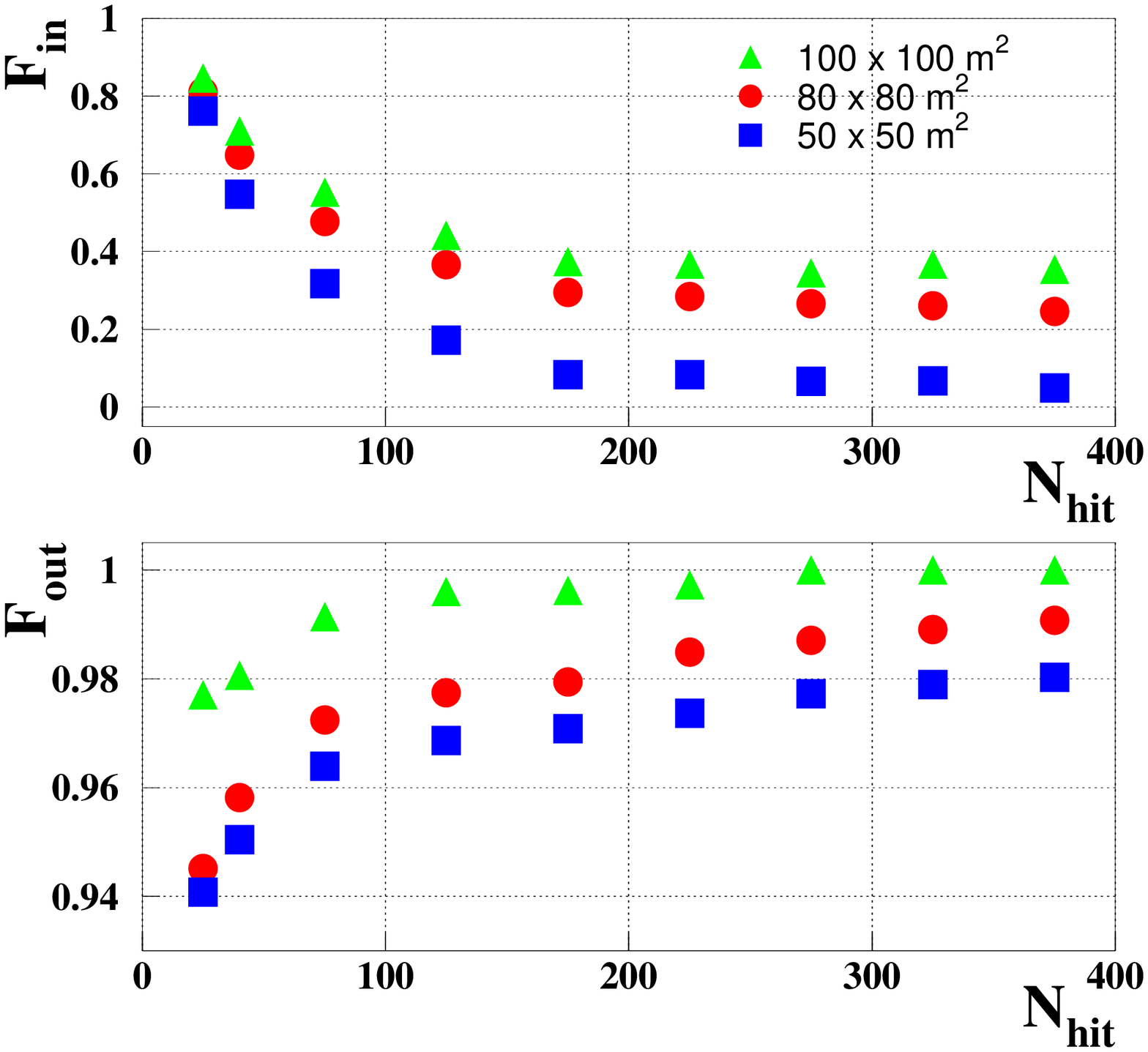}
    \caption{\it Fraction of truly internal ($F_{in}$) and external ($F_{out}$) $\gamma$ 
showers rejected by the selection procedure (1) - (4) for different fiducial areas.
 \label{areasg} }
\end{minipage}\hfill
\begin{minipage}[t]{.48\linewidth}
 \vspace{5.8cm}
    \includegraphics{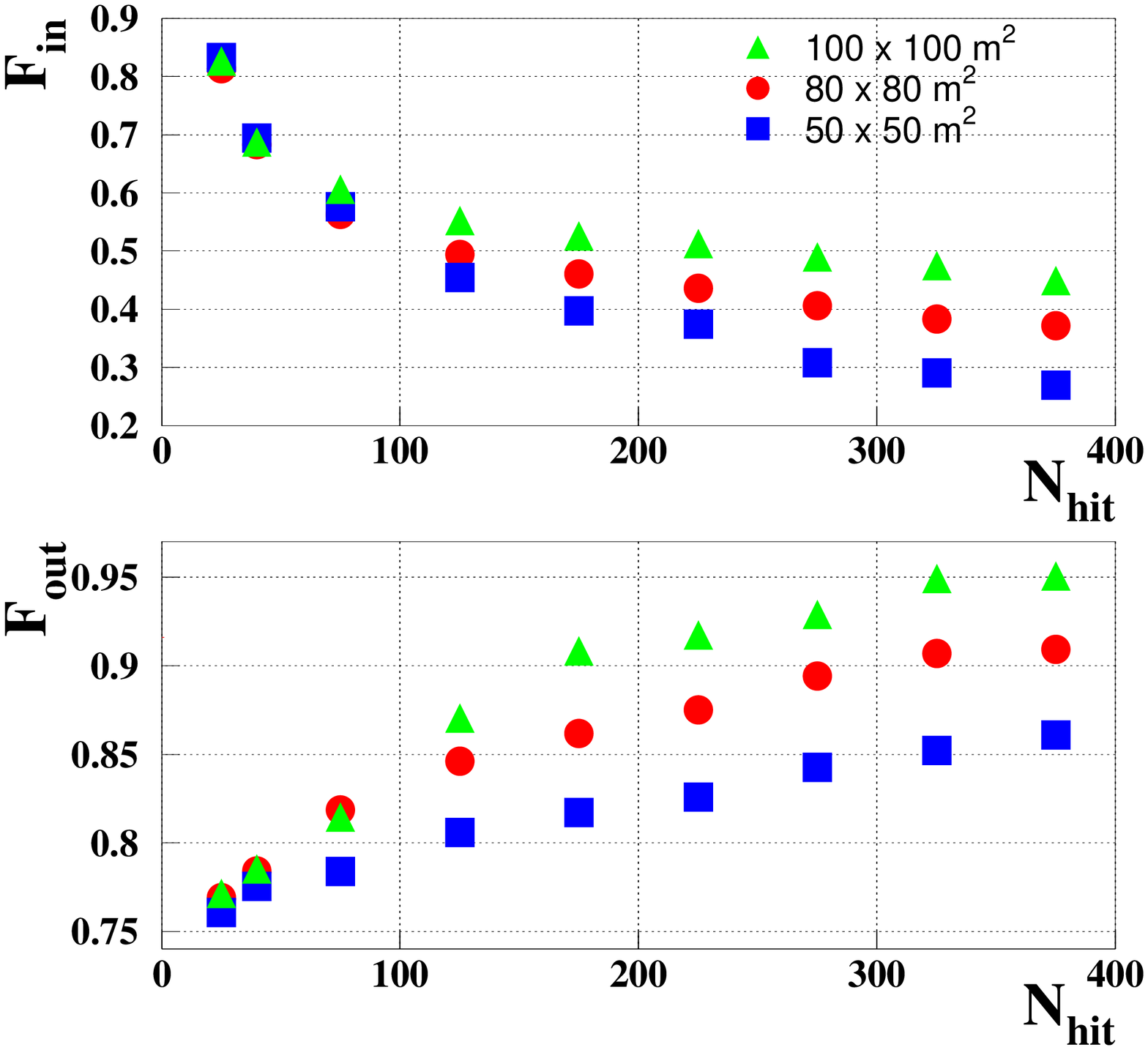}
    \caption{\it Same as Fig. \ref{areasg} but for proton showers.
 \label{areasp} }
\end{minipage}\hfill
\end{figure}

The fraction of internal and external events rejected with the above
procedure is shown for different fiducial areas, separately for 
$\gamma$-induced showers (Fig. \ref{areasg}) and for proton-induced events (Fig. \ref{areasp}).
The fraction of discarded showers increases with $A_{fid}$ mainly at high multiplicities.
For $\gamma$-showers more than $94 \%$ of external events are rejected for the investigated 
fiducial areas. The fraction of internal events erroneously rejected is less than $30 \%$ for high 
multiplicities ($N_{hit} > 150$). The performance of the procedure is lower for proton showers.

Due to the larger lateral particles spread in 
proton showers the $\gamma$/hadron relative trigger efficiency is smaller for 
external events than for internal ones. 
Thus, a suitable choice of the fiducial area $A_{fid}$ may improve the overall detector 
sensitivity to point $\gamma$-ray sources even if the total effective area is decreased.

\section{Conclusions}

The fraction of external events which triggers the ARGO-YBJ detector is relevant, mainly for low
multiplicities (see Fig. \ref{nevents}). 
In this paper we discussed a possible procedure to tag and reject a relevant fraction of external
showers before the calculation of the shower core position via the {\tt LLF2} method, 
thus saving CPU time.
We applied this algorithm to a sample of $\gamma$-induced showers with a Crab-like energy
spectrum. For comparison, we studied the performance of the procedure for proton-induced
showers with an energy spectrum $\sim E^{-2.75}$.

With a suitable choice of the fiducial area and applying simple cuts we can
reject more than $94 \%$ of the external $\gamma$ showers and more than $80 \%$ of the external
proton events, saving more than $50 \%$ of the internal events (for $N_{hit} > 70$).
As a consequence, we expect an improvement in the reconstruction of some shower characteristics, 
such as the primary direction\cite{angul}.



\end{document}